\documentclass[aps,prb,reprint,groupedaddress, showpacs,nofootinbib]{revtex4-2}
\usepackage{graphicx,graphics}
\usepackage{epstopdf}
\usepackage{dcolumn}
\usepackage{amsmath,amssymb,amsfonts}
\usepackage{latexsym,verbatim}
\usepackage{bm}
\usepackage{bbold}
\usepackage{color}
\usepackage{ulem}
\usepackage[breaklinks=true,colorlinks,citecolor=blue,linkcolor=blue,urlcolor=blue]{hyperref}
\usepackage[abs]{overpic}
\usepackage{textcomp}
\usepackage{mathtools}
\usepackage{braket}
\usepackage{soul}

\newif\ifdraft
\drafttrue

\begin{document}

\title{Giant Rabi frequencies between qubit and excited hole states in silicon quantum dots}

\date{\today}

\author{Eleonora Fanucchi$^{1,\S}$}
\author{Gaia Forghieri$^{1,2,\S}$}
\author{Andrea Secchi$^{2}$}
\author{Paolo Bordone$^{1,2}$}
\author{Filippo Troiani$^{2}$}
\affiliation{$^1$Università di Modena e Reggio Emilia, I-41125 Modena, Italy\\ $^2$Centro S3, CNR-Istituto di Nanoscienze, I-41125 Modena, Italy}

\begin{abstract}
Holes in Si quantum dots are being investigated for the implementation of electrically addressable spin qubits. In this perspective, the attention has been focused on the electric-field induced transitions between the eigenstates belonging to the ground doublet. Here we theoretically extend the analysis to the first excited doublet. We show that --- in a prototypical quantum dot structure --- transitions involving the lowest excited states display Rabi frequencies that are several orders of magnitude larger than those occurring in the ground doublet. A clear relation with the symmetries of the eigenstates emerges, {\color{black} as well as} a wide tunability of the Rabi frequencies by means of the applied bias. {\color{black} A preliminary discussion on the} possible implications {\color{black} of the present results} for multilevel manipulation schemes and for multi-hole qubit encodings {\color{black} is provided}.
\end{abstract}

\date{\today}

\maketitle

\section{Introduction}
{\let\thefootnote\relax\footnote{{$^\S$ These authors have equally contributed in this article.}}}
In the last decade, holes in Si and Ge quantum dots (QDs) have been shown to be promising candidates for the implementation of spin qubits \cite{Burkard21a,Scappucci21a,Fang23}.
This essentially results from the strong and tunable spin-orbit (SO) coupling \cite{Kloeffel11,Bosco21a,Bosco21b}, which enables fast all-electric spin manipulation \cite{Watzinger18a,Hendrickx20a,Hendrickx20b,Hendrickx21a,Froning21a,Wang22a,Lawrie23a,Liu23a,Wang24a} and strong coupling with microwave photons \cite{Yu23a}, and from the small hyperfine interactions, leading to long decoherence times \cite{Bosco21c,Piot22a,Hendrickx24a}. 

Hole-spin qubits are implemented in quantum dots, electrostatically defined from planar or one-dimensional Si/Ge heterostructures \cite{Scappucci21a,Fang23}. 
An alternative approach for achieving the three-dimensional confinement of holes is based on metal-oxide semiconductor (MOS) devices \cite{Li15,Maurand16,Crippa2019a,Bonen19,Wang24,Liles24}. This kind of implementation favors integration with the CMOS technology and with the industrial nanofabrication processes, and thus offers potential advantages in terms of scalability \cite{Riggelen21a,Stuyck24a}. 

In the absence of an external magnetic field, single-hole states are organized in Kramers doublets. A static magnetic field $\boldsymbol{B}$ removes the Kramers degeneracy, allowing for an unambiguous definition of the qubit states. These are naturally identified with the two hole states, $\left| e_0 \right>$ and $\left| e_1 \right>$, belonging to the ground doublet. The breaking of time-reversal symmetry induced by $\boldsymbol{B}$, in combination with the strong SO coupling, provides the possibility to manipulate the qubit states electrically \cite{Milivojevic21,Gilbert23,Bosco23}. If the spatial confinement is much stronger in the vertical than in the planar directions, as is typically the case, $\left| e_0 \right>$ and $\left| e_1 \right>$ are (predominantly) heavy-hole states, with marginal light-hole and split-off band components. The possibility of rotating the qubit state entirely depends on such a limited amount of band mixing, and needs to be carefully optimized through the applied magnetic field. In fact, the Rabi frequency is strongly dependent on the orientation of $\boldsymbol{B}$ and --- in typical experimental regimes --- directly proportional to its amplitude \cite{Venitucci18,Venitucci19a,Bellentani21a}. Moreover, {\color{black} if the ground states are characterized by well-defined, albeit not exact, mirror symmetries, the amplitude of transitions induced by spatially homogeneous electric fields tends to be suppressed.}

\begin{figure}
\centering
\includegraphics[width=0.45\textwidth]{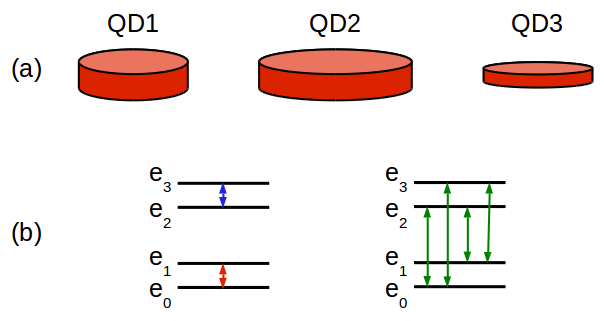}
\caption{\color{black} Schematic representation of the paper contents. (a) We consider three dot geometries: a reference dot geometry (QD1), where we also investigate the effect of variable bias voltage and strain; a dot with a weaker in-plane confinement (QD2), and one with a tighter vertical confinement (QD3). The parameters characterizing each geometry are reported in Table \ref{table1}. (b) For each structure, we consider transitions within the ground (red) and first excited doublets (blue), and interdoblet transitions (green).}
\label{f01}
\end{figure}
{\color{black} Compared to ground states,} excited hole states have comparatively received much less attention \cite{Liles2018}. These states generally present a higher degree of band mixing with respect to those belonging to the ground doublet, and different symmetry properties. In view of the above arguments, the amplitudes of transitions (also referred to as {\it Rabi frequencies}) involving excited states can be orders of magnitude larger than that of the transition between the qubit states $\left| e_0 \right>$ and $\left| e_1 \right>$. Large transition amplitudes represent a potential resource for the {\color{black} hole-based implementation of quantum information. In particular, they suggest the possible use of excited states, either for an alternative encoding of the qubit (or qudit) states, or as auxiliary levels that indirectly couple the conventional qubit states.}

In this work, we compute the Rabi frequencies between the four lowest hole states in a prototypical Si QD. The calculations are based on the Luttinger-Kohn Hamiltonian within the envelope-function formalism \cite{Luttinger55}, a standard approach for the simulation of hole-spin qubits \cite{Venitucci18,Venitucci19a,Bosco21a,Bosco21b,Bosco21c,Secchi21b,Bellentani21a,Forghieri23a,Wang24}. The confinement profile qualitatively reproduces that of Si MOSFETs: in particular, the hole is sharply confined along the growth direction by the band offset between Si and the insulating layers (SiO$_2$), and smoothly confined in the planar directions by the metal-gate induced electrostatic potentials. The gates also generate a bias field along the growth direction, whose interplay with the band profile strongly affects the hole properties and the resulting Rabi frequencies. The main finding of the present investigation is that Rabi frequencies involving hole states from the first excited doublet can exceed that between the ground states by two orders of magnitude. A proper choice of the oscillating-electric field orientation, from the vertical to a specific in-plane direction, leads to further Rabi-frequency enhancements of up to four orders of magnitude, reaching values in the THz range, {\color{black} for static magnetic-field and oscillating electric-field amplitudes of 1 T and 1 mV/nm, respectively}. 

The considered states generally display a significant amount of band mixing. This does not allow for the introduction of effective two-band approaches within the heavy-hole subspace, and the derivation of analytical expressions for the quantities of interest. In order to gain some understanding about the robustness of the observed trends and about the relation between the Rabi frequencies and the main physical parameters that define the QD, we thus report a set of calculations, corresponding to different dot geometries. More specifically, we focus on the relation between the confinement strengths in the vertical and in the in-plane directions, which is known to determine the amount of band mixing in the lowest hole states, and on the interplay between the band offset and the applied bias, which determines the kind of confinement along the growth direction {\color{black} (see Fig.~\ref{f01})}. Overall, this set of calculations shows the robustness of the main findings mentioned above with respect to variations in the dot geometry, and allows us to identify the hole-state symmetry, rather than the amount of band mixing, as a key quantity in determining the Rabi frequencies between states from different doublets. Specifically, the possibility to achieve large Rabi-frequency enhancements depends on the relative symmetry {\color{black}-- along the oscillating-electric-field direction --} of the {\color{black} envelope functions associated with the same spinors in the two states involved in the transition.} The dependence of the Rabi frequencies on (uniaxial compressive) strain that we obtain can also be interpreted along these lines.  

An efficient means for enhancing the Rabi frequencies, including those related to intra-doublet transitions, is represented by the bias field $\boldsymbol{E}_z$. In fact, pronounced maxima in the transition amplitudes are obtained for bias values that correspond to a clear transition in the character of the excited states. Such a transition results in a stark increase of the Rabi frequencies, both for parallel and perpendicular orientations of the oscillating electric field $\delta \boldsymbol{E}$ with respect to $\boldsymbol{E}_z$. 

These results suggest that alternative manipulation schemes, involving the lowest excited states \cite{Vitanov17a,Donarini19,Carter21,Vezvaee23}, might be worth considering {\color{black} in hole-based implementations of quantum-information processing. The involvement of excited states potentially implies further challenges, related, for example, to the shorter decoherence times that typically characterize charge excitations, or to the complexity of controlling the hole state within a multi-level --- rather than a two-level --- computational space. A preliminary discussion of these aspects is presented in the final part of the paper.} Possible implications of the present results also concern multi-hole implementations of the spin qubit. Here, in fact, as an effect of {\color{black} the Pauli principle and of the Coulomb interactions, excited single-hole states enter the composition of all multi-hole states, including those belonging to the qubit space. As a result, the Rabi frequency of the multi-hole qubit results from several contributions, each one corresponding to a single-particle transition amplitude, and most of these transitions involve states that belong to excited doublets.}

The rest of the paper is organized as follows. In Section \ref{sec: method}, we present the theoretical framework and introduce the quantities that are used to characterize the hole states. Section \ref{sec: results} is devoted to the numerical results, and especially to the dependence of the Rabi frequencies on: the confining potential that defines the QD (Subsec.~\ref{subsec:proto}); the applied bias field (Subsec.~\ref{subsec:bias}); the uniaxial compressive strain (Subsec.~\ref{subsec:strain}). {\color{black} In Sec. \ref{sec: manipulation} we discuss various implications of the use of excited states in the qubit (or qudit) manipulation.} Finally, in Section \ref{sec: conclusions} we draw the conclusions. Additional numerical results are reported in the Supplemental Material \cite{SuppMat} and referred to throughout the text. 
\begin{table}[]
    \centering
    \begin{tabular}{|c|cccccc|}
    \hline
       case/dot   & $\hbar\omega_x$ & $\hbar\omega_y$ & $L_z$ & $E_z$ & $-P$ & $\theta$ \\
          & (meV) & (meV) & (nm) & (mV/nm) & (GPa) & (degrees) \\
    \hline
      C1/QD1 & 5 & 6 & 10 & 50 & 0 & $0^\circ$ -- $90^\circ$ \\
      C2/QD2 & {\bf 2.5} & {\bf 3} & 10 & 50 & 0 & $0^\circ$ -- $90^\circ$ \\
      C3/QD3 & 5 & 6 & {\bf 5} & 50 & 0 & $0^\circ$ -- $90^\circ$ \\
      C4/QD1 & 5 & 6 & 10 & {\bf 0 -- 35} & 0 & $75^\circ$ \\
      C5/QD1 & 5 & 6 & 10 & 50 & {\bf 0 -- 1} & $75^\circ$ \\
    \hline
    \end{tabular}
    \caption{Set of parameters that define the considered cases. In all cases, the applied magnetic field has an amplitude of $B=1\,$T, and an orientation defined by the polar ($\theta$) and the azimuthal ($\phi=90^\circ$) angles. The case is defined by the whole set of parameters, the dot only by $\omega_{\alpha=x,y}$ and $L_z$. The bold elements in the table highlight the parameters whose values differ from those of the case C1.}
    \label{table1}
\end{table}

\section{Method}
\label{sec: method}

In the present Section, we briefly outline the theoretical framework (Subsec.~\ref{subsec:hamiltonian}), define the considered quantum-dot model (Subsec.~\ref{subsec:confinement}), and introduce the quantities that are used for characterizing the hole eigenstates (Subsec.~\ref{subsec:manipulation}).

\subsection{Six-band hole Hamiltonian}\label{subsec:hamiltonian}
In Si, the multiband character of the hole eigenstates plays a crucial role in determining the relevant qubit properties, especially the intradoublet energy (Zeeman) splittings and the Rabi frequencies. In fact, these result from the interplay between the atomic spin-orbit coupling, the confining potential, and the external magnetic field. In order to account for these features, we compute the hole states by diagonalizing the six-band Luttinger-Kohn Hamiltonian \cite{Luttinger55, Voon09}, with the inclusion of an external confining potential that defines the quantum-dot model. 

Within this formalism, the $k$-th hole eigenstate, whose energy is denoted by $e_k$ (with $k=0,1,\dots$), is written in the position representation as
\begin{align}
    \left| e_k \right> = \sum_{J,M} \psi_{k; J,M}(\boldsymbol{r})  \left| J, M \right> \,.
\end{align}
Here, $\left| J, M \right> $ is a band spinor, codifying a Bloch state and multiplying  the envelope function $\psi_{k; J,M}(\boldsymbol{r})$. 
In Si, the top of the valence band typically requires the inclusion of $J \in \left\{ \frac{3}{2}, \frac{1}{2} \right\}$ and $M \in \left\{ -J, -J+1, \ldots, J \right\}$. Spinors with $J = \frac{3}{2}$ and $|M| = \frac{3}{2}$ or $|M| = \frac{1}{2}$ correspond, respectively, to the heavy-hole (HH) and to the light-hole (LH) components; spinors with $J = \frac{1}{2}$ correspond to the spin-orbit split-off band of holes (SH). 

The Hamiltonian, whose diagonalization determines the eigenenergies and the envelope functions, is thus a $6 \times 6$ matrix in the spinor basis; its elements are functions of the hole position $\boldsymbol{r}$ and of the momentum operator $\boldsymbol{k}$. It can be written as the sum of four terms: 
\begin{align}
    H =  H_{\boldsymbol{k} \cdot \boldsymbol{p}}   +  V  +  H_{\rm magn}  +  H_{\rm strain}  \,.
    \label{total H}
\end{align}
\begin{widetext}
Here, if the basis is ordered as $\left\{ \left| \frac{3}{2}, \frac{3}{2}\right> , \left| \frac{3}{2}, \frac{1}{2}\right> , \left| \frac{3}{2}, -\frac{1}{2}\right> , \left| \frac{3}{2}, -\frac{3}{2}\right> , \left| \frac{1}{2}, \frac{1}{2}\right> , \left| \frac{1}{2}, -\frac{1}{2}\right> \right \} $, the kinetic term reads as
\begin{eqnarray}
    H_{\boldsymbol{k} \cdot \boldsymbol{p}}   = 
    \left(
    \begin{array}{cccccc}
              P+Q &             -S &             R &             0 &    -S\tfrac{1}{\sqrt{2}} & R \sqrt{2}     \\
             -S^* &            P-Q &             0 &             R &    -Q \sqrt{2} & S \tfrac{\sqrt{3}}{\sqrt{2}}   \\
              R^* &              0 &           P-Q &             S & S^* \tfrac{\sqrt{3}}{\sqrt{2}}  & Q\sqrt{2} \\
                0 &            R^* &           S^* & P+Q           &  -R^* \sqrt{2} & -S^*\tfrac{1}{\sqrt{2}}  \\
    -S^*\tfrac{1}{\sqrt{2}} &  -Q^* \sqrt{2} & S \tfrac{\sqrt{3}}{\sqrt{2}} &    -R \sqrt{2} &  P+\Delta_{\rm SO} &              0 \\
     R^* \sqrt{2} & S^* \tfrac{\sqrt{3}}{\sqrt{2}}  &  Q^*\sqrt{2} &   -S\tfrac{1}{\sqrt{2}} &              0 &  P+\Delta_{\rm SO} \\
    \end{array}
    \right),
   \label{k.p Hamiltonian}
\end{eqnarray}
\end{widetext}
where $\Delta_{\rm SO}=44\,$meV is the spin-orbit parameter in Si. The operators appearing in the above matrix are:
\begin{align}
   & P  =  \frac{\hbar^2}{2m_0}\gamma_1 (k_x^2+k_y^2+k_z^2) \,,  \nonumber \\ 
   & Q  = \frac{\hbar^2}{2m_0}\gamma_2 (k_x^2+k_y^2-2k_z^2) \,,  \nonumber \\ 
   & R  = \frac{\hbar^2}{2m_0}\sqrt{3} [-\gamma_3(k_x^2-k_y^2)+2i\gamma_2 k_x k_y] \,, \nonumber \\   
   & S  = \frac{\hbar^2}{2m_0}2\sqrt{3} \gamma_3(k_x-ik_y)k_z \,, 
\end{align}
where $m_0$ is the electron mass. The Luttinger parameters for Si are: $\gamma_1  = 4.285$, $\gamma_2  =  0.339$, and $ \gamma_3 = 1.446$. The expressions of these operators, as well as the form of Eq.~\eqref{k.p Hamiltonian}, is based on the following correspondence between the coordinate and crystallographic axes \cite{Venitucci18, Secchi21a}:
\begin{align}
    \boldsymbol{u}_x \parallel [110] \,, \quad \boldsymbol{u}_y \parallel [\overline{1}10] \,, \quad \boldsymbol{u}_z \parallel [001] \,.
\end{align}

The second term in Eq.~\eqref{total H}, $V$, is the confining potential, assumed to have a slow spatial dependence with respect to the lattice periodicity and, therefore, to leave the Bloch states unmixed; formally, it is given by a function of position $V(\boldsymbol{r})$, multiplied by the $6 \times 6$ identity matrix in the band-spinor space. 
The third term, $H_{\rm magn}$, groups all the contributions that depend on the magnetic field $\boldsymbol{B}$, including the Zeeman-Bloch term \cite{Venitucci18}. The other terms, which are linear and quadratic in the components of $\boldsymbol{B}$, are obtained by applying the Peierls substitution to $H_{\boldsymbol{k} \cdot \boldsymbol{p}}$, where the vector potential is given by $\boldsymbol{A} = - \frac{1}{2} \boldsymbol{r} \times \boldsymbol{B}$ \cite{Bellentani21a}. 
Finally, $H_{\rm strain}$ is the Bir-Pikus Hamiltonian that accounts for homogeneous strain \cite{BirPikus, Chao92a}. In the present paper, we consider prototypical QDs either in the absence of strain, or in the presence of a uniaxial compressive strain along the $[110]$ crystallographic direction, which coincides with the $x$ axis in the chosen reference frame. This strain is typically applied to the Si channels in advanced MOSFETs in order to increase the hole mobility \cite{Sverdlov}.

\subsection{Quantum-dot model}\label{subsec:confinement}

We consider a prototypical QD model, defined by a parabolic confinement in the planar ($x$ and $y$) directions and by a biased square well in the vertical direction ($z$). The use of this potential as a good low-energy approximation of real systems is supported both by experimental evidence \cite{Liles2018} and by numerical simulations of the confinement profiles, e.g., in industrial MOSFET devices \cite{Bellentani21a}. The overall confinement potential is thus given by the sum of three terms, each of which depends on a different coordinate: $V(\boldsymbol{r}) = V_x (x) + V_y(y) + V_z(z)$, where:
\begin{gather}\label{eq:confxy}
    V_{x} (x) = \frac{1}{2} \kappa_x x^2,\quad V_{y} (y) = \frac{1}{2} \kappa_y y^2\,,
\end{gather}
and 
\begin{gather}
    V_{z} (z) = \begin{cases}
        -E_z z\,, & {\rm for}\ \left|z\right|<L_z/2 \\
        V_{\rm bo}\,, & {\rm for}\ \left|z\right|>L_z/2
    \end{cases}\, .
\end{gather}
Here, $V_{\rm bo}$ is the band-offset between the semiconductor (Si) and the surrounding insulator materials, $L_z$ is the well width, and $E_z$ is the bias field applied along the vertical direction. We take $V_{\rm bo} = 4$ eV, a typical value for the Si/SiO$_2$ interface.

The parameters $\kappa_{\alpha}$, which determine the strength of the confinement in the $\alpha = x,y$ directions [Eq.~\eqref{eq:confxy}], can be related to the characteristic in-plane excitation energies $\hbar\omega_{\alpha} \equiv \hbar(\gamma_1\kappa_{\alpha}/m_0)^{1/2}$ and confinement lengths $l_{\alpha} \equiv (\hbar\gamma_1/m_0\omega_{\alpha})^{1/2}$. 
The spatial confinement along the vertical direction is characterized by two independent length scales: the well width $L_z$ and $l_E \equiv (\hbar^2\gamma_1/2m_0E_z)^{1/3} $, associated with the bias field $E_z$ \cite{Bosco21a}. The values of the parameters entering the confinement potentials that have been used in the calculations are reported in Table \ref{table1} and in the table caption. 

\subsection{\color{black} Larmor and Rabi frequencies}\label{subsec:larmor_rabi}

Within the standard approach to the hole-spin qubit implementation, the logical states $|0\rangle$ and $|1\rangle$ are identified with the hole eigenstates belonging to the ground doublet, $|e_0\rangle$ and $|e_1\rangle$, whose Kramers degeneracy is removed by the applied magnetic field. The amplitude of the transition between these two states, induced by an oscillating electric field $\delta\boldsymbol{E}$, defines (up to a factor $h$) the Rabi frequency. In this paper we consider not only the ground doublet, but also the two eigenstates $|e_2\rangle$ and $|e_3\rangle$, belonging to the first excited doublet{\color{black}, which can be involved in the encoding and manipulation of quantum information within different approaches (see Section \ref{sec: manipulation})}.

For each pair of eigenstates, the energy gap is
\begin{gather}
    \Delta_{jk} \equiv e_j - e_k
\end{gather}
and the Rabi frequency associated with the time-dependent manipulation potential $ \delta V_{\alpha}(t) = \alpha\, \delta E_\alpha \cos (\omega t)$ is given by
\begin{equation}\label{eq:rabi_x}
    f_{\alpha; jk} \equiv \frac{1}{h}  \big| \delta E_{\alpha} \left\langle e_j|\hat{\alpha}|e_k\right\rangle \big| \, .
\end{equation}
Here, $\alpha = x,y,z $ is the component of the hole position operator along the direction parallel to the applied oscillating electric field. In the following, we consider field orientations along the weakest ($\alpha = x$) and strongest ($\alpha = z$) confinement directions.   

\subsection{Characterization of the hole eigenstates}\label{subsec:manipulation}

While the Rabi frequencies are the main point of interest in the present investigation, other quantities are also used to characterize the hole eigenstates and relate the tunable system parameters to the values of the transition amplitudes.
These auxiliary quantities include the band occupations
\begin{align}
    w_{k;J,|M|} \equiv \sum_{\chi = \pm 1} \int d \boldsymbol{r}  \left| \psi_{k; J, \chi M}(\boldsymbol{r}) \right|^2  \,,
\end{align}
corresponding to the heavy-hole ($J\!=\!|M|\!= \! 3/2$), light-hole ($J \!= \! 3/2$, and $|M| \! = \! 1/2$), and split-off ($J\!=\! |M| \!=\!1/2$) bands for the state $|e_k\rangle$.  

\begin{figure*}
\centering
\includegraphics[width=\textwidth]{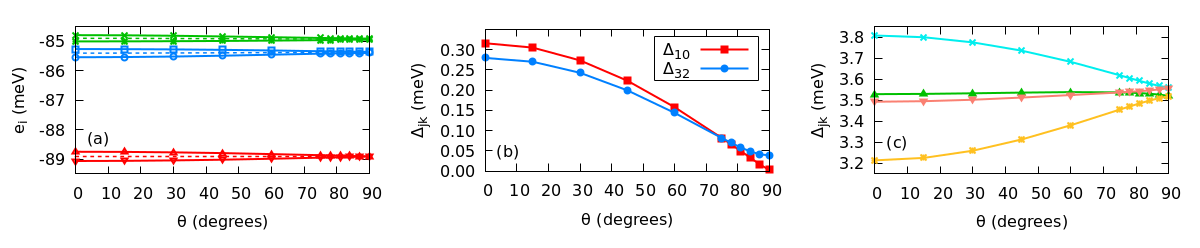}
\caption{(a) Lowest energy levels $e_i$ ($i=0,\dots,3$) and (b,c) corresponding energy gaps $\Delta_{jk} \equiv e_j - e_k$ for holes in the reference quantum dot QD1. The color code for the energy gaps $\Delta_{jk}$ between states belonging to different doublets, displayed in panel (c), is the following: $\Delta_{21}$ (yellow), $\Delta_{31}$ (salmon), $\Delta_{20}$ (green), $\Delta_{30}$ (cyan). {\color{black} The values of $\omega_x$, $\omega_y$, $L_z$, $E_z$, and $B$ used in these calculations are reported in Table \ref{table1} (case C1) and in the table caption.} 
}
\label{f02}
\end{figure*}
Besides, we consider the spatial parities along the coordinate axes. These are quantified through the expectation values {\color{black} of the mirror reflection operators $\sigma_{\alpha\beta}$, whose effect is that of reflecting the envelope functions about the plane $\alpha\beta = xy, yz, zx$. The expressions of such expectation values read:} 
\begin{align}
    & \left< e_k \right| \hat{\sigma}_{xy} \left| e_k \right> = \sum_{J,M} \int d \boldsymbol{r}\,   \psi^*_{k; J, M}(x,y,z) \,  \psi_{k; J, M}(x,y,-z)  \,, \nonumber \\
    & \left< e_k \right| \hat{\sigma}_{zx} \left| e_k \right> = \sum_{J,M} \int d \boldsymbol{r}\,   \psi^*_{k; J, M}(x,y,z) \,  \psi_{k; J, M}(x,-y,z)  \,, \nonumber \\
    & \left< e_k \right| \hat{\sigma}_{yz} \left| e_k \right> = \sum_{J,M} \int d \boldsymbol{r}\,   \psi^*_{k; J, M}(x,y,z) \,  \psi_{k; J, M}(-x,y,z)  \,.
    \label{parity exp values}
\end{align}
{\color{black} The first of the above equations results from the fact that $\hat\sigma_{xy}  \psi_{k; J, M} (x,y,z) = \psi_{k; J, M} (x,y,-z)$; analogous identities hold for the other reflection planes. Hole eigenstates that are exactly symmetric or antisymmetric with respect to the $xy$ plane are characterized by the property $\psi_{k; J, M} (x,y,-z) = \pm\psi_{k; J, M} (x,y,z) $ for all $J$ and $M$, and therefore by the expectation values $\left< e_k \right| \hat{\sigma}_{xy} \left| e_k \right> = \pm 1$. A state with $\langle\hat{\sigma}_{\alpha\beta}\rangle=0$ is characterized instead by an undefined $xy$ mirror symmetry, i.e. by the sum of a symmetric and an antisymmetric component of equal weight. Analogous considerations hold for the operators $\hat\sigma_{yz}$ and $\hat\sigma_{zx}$. 

In the cases considered hereafter, the hole eigenstates do not possess exact mirror symmetries, because} none of the mirror reflection operators commutes with the Hamiltonian. {\color{black} Besides, different band components $\psi_{k; J, M} (\boldsymbol{r})$ of a given state $|e_k\rangle$ can be characterized by different symmetries.  However, the expectation values $\langle\hat{\sigma}_{\alpha\beta}\rangle$ represent a useful means to characterize the hole eigenstates, and capture most of the transitions these undergo as a function of the external parameters (magnetic-field orientation, bias field, strain).}

Finally, the two states belonging to each doublet, whose energies are split by the magnetic field, are identified by means of the {\it pseudospin}. In particular, we assign to the states whose energy is lowered (increased) by a magnetic field with a given orientation the pseudospin value $-1$ ($+1$). {\color{black} Two states that are distinguished by the pseudospin tend to form a degenerate Kramers doublet when $\boldsymbol{B} \rightarrow \boldsymbol{0}$; therefore, they are close to being one the time-reversal conjugate of the other, and to having the same expectation values of the mirror symmetries.}

\section{Results}
\label{sec: results}

\begin{figure*}
\centering
\includegraphics[width=\textwidth]{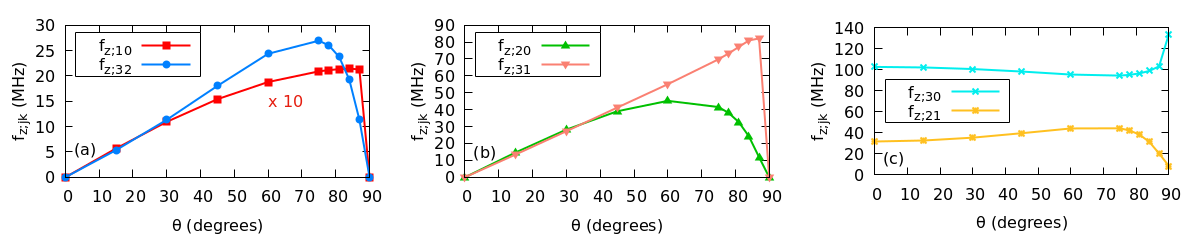}
\includegraphics[width=\textwidth]{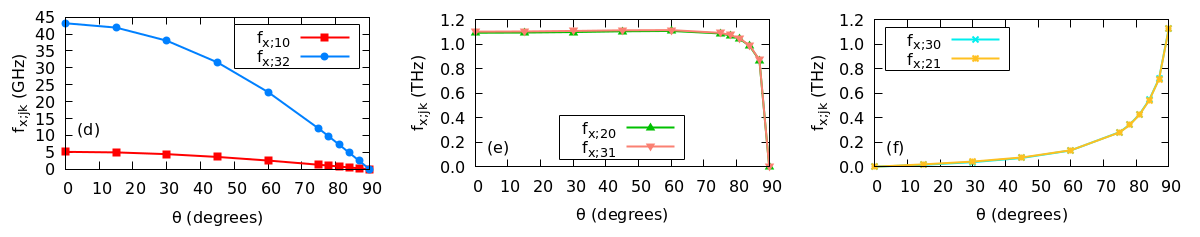}
\caption{Rabi frequencies (a,b,c) $f_{z;jk}$ and (d,e,f) $f_{x;jk}$ for the dot QD1, associated to a time-dependent field of amplitude $\delta E=1\,$mV/nm, oriented along the $z$ and $x$ directions, respectively.  {\color{black} The values of $\omega_x$, $\omega_y$, $L_z$, $E_z$, and $B$ used in these calculations are reported in Table \ref{table1} (case C1) and in the table caption.}
}
\label{f03}
\end{figure*}
The numerical results presented hereafter refer to different {\color{black} dot geometries [Fig. \ref{f01}(a) and Table \ref{table1}] and to the variation of different tunable parameters. The comparison between these cases is} meant to highlight the dependence of the dot properties, and specifically that of the Rabi frequencies, on the parameters that define the {\color{black} in-plane and vertical confinements} (Subsec.~\ref{subsec:proto}), on the bias field (Subsec.~\ref{subsec:bias}), and on the compressive uniaxial strain (Subsec.~\ref{subsec:strain}). 

\subsection{Prototypical quantum dots}
\label{subsec:proto}
The geometry of the dot is here defined by the values of $\omega_x$ and $\omega_y$ on the one hand, and by that of $L_z$ (well width) on the other. The first set of calculations presented hereafter concerns a reference quantum dot, referred to as QD1, which corresponds to the case C1 in Table \ref{table1} (Subsubsec.~\ref{subsubsec:rqd}). In order to investigate the role of the dot geometry, and specifically the dependence of the Rabi frequencies on the relative strength of the in-plane and vertical confinements, the results obtained for QD1 are compared with those obtained for two other dots, referred to as QD2 (C2, Subsubsec.~\ref{subsubsec:wipc}) and QD3 (C3, Subsubsec.~\ref{subsubsec:svc}), characterized by a weaker in-plane confinement and by a stronger vertical confinement, respectively {\color{black} [Fig. \ref{f01}(a) and Table \ref{table1}]}. 

\subsubsection{Reference quantum dot}
\label{subsubsec:rqd}

The first considered case (C1) corresponds to the reference quantum dot (QD1). The values of the relevant physical parameters are reported in Table \ref{table1}. The characteristic length scales of the in-plane confinement for QD1 are: $l_x\approx 8.08\,$nm and $l_y\approx 7.38\,$nm.

\paragraph{Energy spectrum.}
As a first means to characterize the low-energy properties of the dot, we consider the lowest energy eigenvalues $e_i$ ($i=0,\dots,3$) of the LK Hamiltonian, and the corresponding energy gaps $\Delta_{jk}\equiv e_j -e_k$ (Fig.~\ref{f02}). The energy spectrum is characterized by the presence of a well-defined ground doublet [red symbols, panel (a)] and of two excited doublets (blue and green). These can be approximately related to excitations along the $x$ and $y$ directions, as witnessed by inspection of the main eigenstate components.

The intra-doublet energy gaps [panel (b)], $\Delta_{10}$ and $\Delta_{32}$, display a similar, monotonically decreasing dependence on the polar angle $\theta$, which specifies the orientation of the magnetic field with respect to the $z$ axis. The inter-doublet energy gaps [panel (c)] are either insensitive to the field orientation or vary monotonically with $\theta$, depending on whether they involve states with equal ($\Delta_{20}$ and $\Delta_{31}$) or opposite ($\Delta_{21}$ and $\Delta_{30}$) pseudospins. 

\paragraph{Hole eigenstates.}
The hole eigenstates can be characterized through their mirror symmetries. Even though the Hamiltonian does not possess exact mirror symmetries, the lowest eigenstates display an approximately symmetric or antisymmetric character with respect to (some of) the coordinate planes. This can be related to the fact that $V_x(x)$ and $V_y(y)$ are even under the transformations $x \rightarrow -x$ (induced by $\hat{\sigma}_{yz}$) and $y \rightarrow -y$ (induced by $\hat{\sigma}_{zx}$), while $V_z(z)$ has no defined $xy$ mirror symmetry. As a result, the ground states ($k=0,1$) are approximately symmetric with respect to the $yz$ and $zx$ planes for all values of $\theta$:
\begin{gather}
\langle e_k | \hat{\sigma}_{yz} | e_k \rangle \approx \langle e_k | \hat{\sigma}_{zx} | e_k \rangle \approx  0.99  
\end{gather}
while their $xy$ mirror symmetry is undefined 
\begin{gather}
\langle e_k | \hat{\sigma}_{xy} | e_k \rangle \approx 0.045\,.
\end{gather} 
The states belonging to the first excited doublet are approximately antisymmetric along the $x$ direction 
\begin{gather}
\langle e_2 | \hat{\sigma}_{yz} | e_2 \rangle\approx -0.76 \,, \quad  
\langle e_3 | \hat{\sigma}_{yz} | e_3 \rangle\approx -0.82\,.
\end{gather}
\begin{figure*}
\centering
\includegraphics[width=\textwidth]{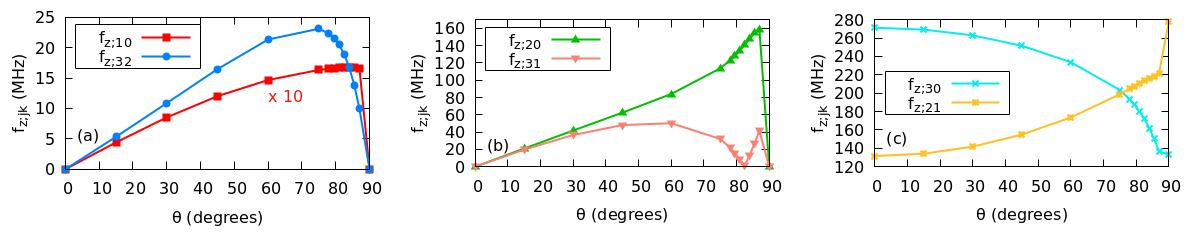}
\caption{(a,b,c) Rabi frequencies $f_{z;jk}$ for the dot QD3, associated to a time-dependent field of amplitude $\delta E=1\,$mV/nm, oriented along the $z$ and $x$ directions, respectively.  {\color{black} The values of $\omega_x$, $\omega_y$, $L_z$, $E_z$, and $B$ used in these calculations are reported in Table \ref{table1} (case C3) and in the table caption.}}
\label{f04}
\end{figure*}
This confirms that the lowest excitation is mainly related to the motion along the direction with the weakest spatial confinement (which is $x$, being $l_x > l_y > l_E $).
{\color{black} Further details on the symmetries of the hole states are provided in the Supplemental Material (Fig.~S8) \cite{SuppMat}}. 

As a second means to characterize the hole eigenstates, we consider the occupation of the bands. The states belonging to the ground doublet are predominantly HH ($w_{k;3/2,3/2} \approx 0.98$, for $k=0,1$), with minor contributions from the LH and SH bands. The states belonging to the first excited doublet display a larger degree of band mixing ($w_{k;3/2,3/2} \approx 0.91 $, for $k=2,3$). The lowest excitation observed in the hole states thus has a hybrid character, which involves both the motion along the $x$ direction and, to a minor extent, the band composition (see also Fig.~S3 in the Supp.~Mat.~\cite{SuppMat}). 

\paragraph{Rabi frequencies.}
The main results of the present Subsubsection are the values of the Rabi frequencies. We are specifically interested in two aspects, which are independent of the values adopted for the fields in the present calculations {\color{black} ($B=1\,$T and $\delta E=1\,$mV/nm)}: $(i)$ the ratio between the Rabi frequencies involving only the states of the ground doublet and those that involve the excited states; $(ii)$ the dependence of the Rabi frequencies on the orientation of the oscillating electric field. {\color{black} Being the Rabi frequencies linear both in $B$ and $\delta E$, their values for different field intensities can be obtained from the current ones through a proportional rescaling}. 

We start by considering the case in which the oscillating electric field is parallel to the growth direction $z$ (Fig.~\ref{f03}). The Rabi frequency involving the excited states ($f_{z;32}$) is on average one order of magnitude larger than that involving the ground states ($f_{z;10}$), and displays a similar dependence on $\theta$ [panel (a)]. Inter-doublet transitions are characterized by even larger Rabi frequencies, of the order of $10^2$ MHz, both in the case of pseudospin-preserving [$f_{z;20}$ and $f_{z;31}$, panel (b)] and in that of pseudospin-flipping transitions [$f_{z;30}$ and $f_{z;21}$, panel (c)]. 

We then consider the case where the oscillating electric field $\delta\boldsymbol{E}$ is oriented along the $x$ direction [panels (d,e,f)]. {\color{black} Here, the Rabi frequencies related to intradoublet transitions are three orders of magnitude larger than the corresponding $f_{x; jk}$, while those related to interdoublet transitions achieve the THz regime.} Finally, we note that the dependence of the Rabi frequencies $f_{x;jk}$ on $\theta$ is qualitatively different from that obtained when the field is parallel to $z$: this includes the presence of maxima at $\theta=0$, where $f_{z;jk}$ typically displays minima. 

\subsubsection{Weaker in-plane confinement\label{subsubsec:wipc}}

The second considered case (C2, second line in Table \ref{table1}) corresponds to the quantum dot QD2. This presents values of $\omega_x$ and $\omega_y$ that are halved with respect to those of QD1, resulting in larger length scales: $l_x\approx 11.4\,$nm and $l_y\approx 10.4\,$nm. 

\paragraph{Energy spectrum.}
Both the level structure and the symmetries of the eigenstates closely resemble those obtained for QD1 (see Figs.~S1 and S9 in the Supp. Mat. \cite{SuppMat}). 

\paragraph{Hole eigenstates.}
The reduced strength of the in-plane confinement results --- as one might expect --- in a suppressed band mixing, i.e. in a reduced occupation of the LH and SH bands. This applies to both the ground and the first-excited doublets, the latter one displaying a stronger dependence on the magnetic-field orientation  (see Fig.~S4 in the Supp.~Mat.~\cite{SuppMat}). 

\paragraph{Rabi frequencies.}
The Rabi frequencies $f_{\alpha;jk}$ obtained for QD2 (Fig.~S13 in the Supp. Mat. \cite{SuppMat}) display behaviors that are very similar to those of QD1.
In absolute terms, the $f_{z;jk}$ frequencies are reduced, particularly that between the states of the ground doublet.
Instead, a systematic but moderate increase is observed in the $f_{x;jk}$ frequencies.

Overall, these results indicate a robustness of the trends with respect to the details of the confinement potential. They also suggest that the degree of band mixing alone cannot be regarded as a decisive factor. In fact, while the band mixing clearly decreases in all the considered eigenstates while passing from QD1 to QD2, this does not result in a general reduction of the Rabi frequencies. 

\subsubsection{Stronger vertical confinement\label{subsubsec:svc}}

The third considered case (C3, third line in Table \ref{table1}) corresponds to the quantum dot QD3. This is characterized by a stronger vertical confinement with respect to the reference dot, resulting from the reduction of the well width $L_z$ by a factor 2. 

\begin{figure*}[t]
\centering
\includegraphics[width=\textwidth]{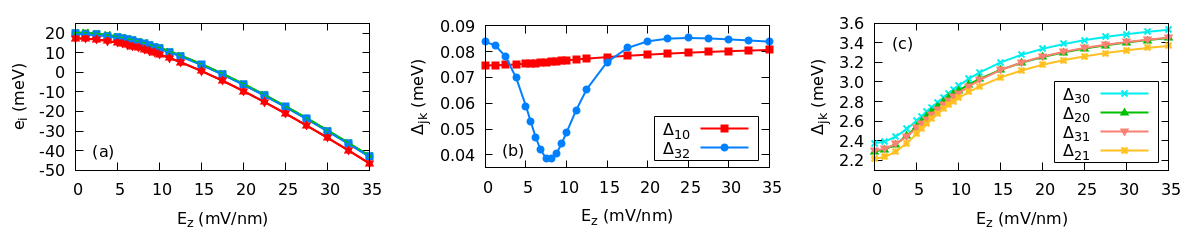}
\caption{(a) Lowest energy levels $e_i$ and (b,c) corresponding energy gaps $\Delta_{jk} \equiv e_j - e_k$ as a function of the bias field $E_z$, for holes in the QD1.  {\color{black} The values of $\omega_x$, $\omega_y$, $L_z$, $\theta$, and $B$ used in these calculations are reported in Table \ref{table1} (case C4) and in the table caption.}}
\label{f05}
\end{figure*}
\paragraph{Energy spectrum.}
Since the lowest excitation energies are related to the in-plane confinement, the energy levels $e_i$ and (thus) the energy gaps $\Delta_{jk}$ are essentially unchanged, with respect to those obtained for QD1 (see Fig.~S2 in the Supp. Mat. \cite{SuppMat}).

\paragraph{Hole eigenstates.}
The reduced well width affects the properties of the eigenstates. In particular, their $xy$ mirror symmetry is more defined than in QD1. This applies both to the ground  and to the first excited doublets:
\begin{gather}
\langle e_{j=0,1} | \hat{\sigma}_{xy} | e_j \rangle\approx 0.81\,,\quad 
\langle e_{k=2,3} | \hat{\sigma}_{xy} | e_k \rangle\approx 0.79\,. 
\end{gather}
The symmetries along the two other coordinate directions are scarcely (ground doublet) or moderately (excited doublet) affected by the reduction of the well width (see Fig.~S10 in the Supp. Mat. \cite{SuppMat}). 

The increased strength of the vertical confinement also results in a reduced band mixing with respect to QD1. In particular, the overall occupation of the LH subbands is below 1.5\% and around 5\% for the ground and first excited doublets, respectively (see also Fig.~S5 in the Supp. Mat. \cite{SuppMat}). The difference with respect to the band occupation of QD1 is lower than that  obtained for QD2, even though the reduction of the well width by a factor 2 should lower the relative strength of the in- and out-of-plane confinements more than a reduction by a factor 2 ($\sqrt{2}$) of $\omega_x$ and $\omega_y$ ($l_x$ and $l_y$). This apparent anomaly is due to the fact that the vertical confinement in the considered QDs is determined not only by the well width, but also by the bias. Therefore, the difference between the actual strength of the vertical confinement in QD1 and QD3 is smaller than what suggested simply by the ratio between the respective values of $L_z$. 

\paragraph{Rabi frequencies.}
The stark changes in the mirror symmetries of the hole eigenstates are reflected in the Rabi frequencies and in their dependence on the magnetic-field orientation. In particular, while the Rabi frequencies $f_{z;jk}$ related to the intra-doublet transitions hardly differ from those of QD1 [Fig.~\ref{f04}(a)], those related to interdoublet transitions display quantitative and qualitatively differences, with maxima that are more than doubled [panels (b,c)]. 

The Rabi frequencies $f_{x;jk}$ (see Fig.~S14 in the Supp. Mat. \cite{SuppMat}), instead, do not present significant differences with respect to those obtained for QD1. This reflects the scarce effect of the reduction of the well width on the symmetries in the planar directions.

\subsubsection{General remarks on the role of dot geometry}
A few remarks can be drawn from the comparison between the Rabi frequencies obtained for the three considered quantum-dot structures (QD1, QD2, and QD3). Concerning the dependence on the hole eigenstates, transitions within the excited doublet ($f_{\alpha;32}$, with $\alpha = x,z$) allow for an increase of one order of magnitude with respect to those within the ground doublet ($f_{\alpha;10}$). From one to two further orders of magnitude can be gained by moving to interdoublet transitions ($f_{\alpha;jk}$, with $j=0,1$ and $k=2,3$). Concerning the dependence on the orientation of the oscillating field $\delta\boldsymbol{E}$, the Rabi frequencies $f_{x;jk}$ are from three to four orders of magnitude larger than the $f_{z;jk}$ involving the same pairs of eigenstates. 

These features are common to all three dots, in spite of the significant differences between the values of the parameters that determine the confining potential. As to the differences between the quantum dots, and thus to the relation between the transition amplitudes and quantities that characterize the hole eigenstates, a clear connection with the degree of band mixing in the eigenstates emerges only for the amplitudes of intradoublet transitions $f_{z;10}$ and $f_{z;32}$. The large values of the Rabi frequencies obtained for in-plane electric fields ($f_{x;jk}$) can be connected instead to the fact that both the involved hole eigenstates and the potential related to the oscillating electric field have well-defined symmetries along the same direction. In particular, the largest (smallest) $f_{\alpha;jk}$ are obtained for hole states $|e_j\rangle$ and $|e_k\rangle$ characterized by opposite (equal) symmetries along the direction of the homogeneous field $\delta\boldsymbol{E}$. Further insight in this respect is provided by the dependence of the Rabi frequencies on the bias field and on the strain (see below).

\subsection{Effect of the bias field\label{subsec:vb}}
\label{subsec:bias}

\begin{figure*}[t]
\centering
\includegraphics[width=\textwidth]{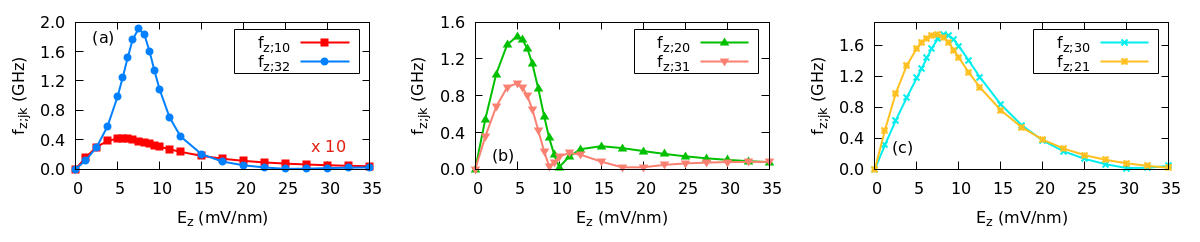}
\includegraphics[width=\textwidth]{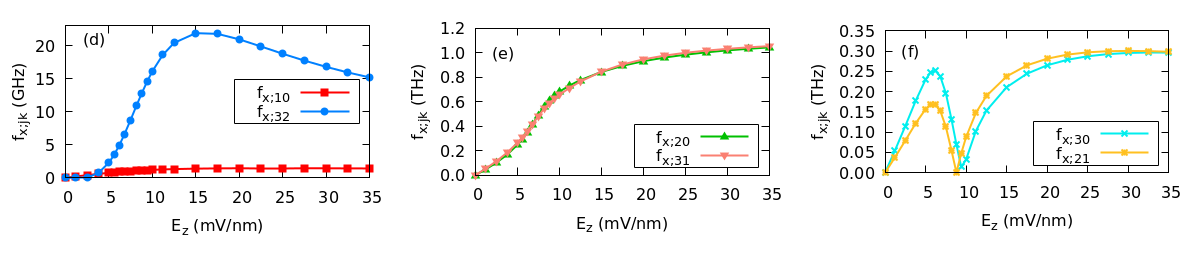}
\caption{Rabi frequencies (a,b,c) $f_{z;jk}$ and (d,e,f) $f_{x;jk}$ for QD1 as a function of the applied bias $E_z$, associated to a time-dependent field of amplitude $\delta E=1\,$mV/nm, oriented along the $z$ and $x$ directions, respectively.  {\color{black} The values of $\omega_x$, $\omega_y$, $L_z$, $\theta$, and $B$ used in these calculations are reported in Table \ref{table1} (case C4) and in the table caption.}}
\label{f06}
\end{figure*}
The interplay between the (anti-symmetric) bias-induced potential and the (symmetric) square-well component of $V_z(z)$ strongly affects the properties of the hole eigenstates. In order to further investigate such interplay, we consider the dependence on the bias field $E_z$ of the quantities considered so far in QD1, for a fixed orientation of the magnetic field (see C4 in Table \ref{table1}).  

\paragraph{Hole energies and eigenstates.}
All of the considered energies $e_i$ display a nonlinear and comparable Stark shift as a function of $E_z$ [Fig.~\ref{f05}(a)]. The dependence on the bias of the energy splitting within the excited doublet, $\Delta_{32}$, shows a clear minimum for a value of $E_z$ in the interval $5 \div 15\,$mV/nm [panel (b)], which indicates the occurrence of an avoided level crossing and of a related transition in the character of the states $|e_2\rangle$ and $|e_3\rangle$ (see below). The gap between the doublets increases monotonically with $E_z$: a robust trend that does not appear to be affected by the above mentioned transition [panel (c)].

The effect of the bias on the hole eigenstates clearly emerges from the expectation values of the mirror symmetry operators $\hat{\sigma}_{\alpha \beta}$. All four states ($k=0,1,2,3$) undergo a clear transition in the $xy$ mirror symmetry, going from a symmetric ($  \langle e_k | \hat\sigma_{xy} | e_k \rangle\!\approx\! 1$) to an undefined ($\langle e_k | \sigma_{xy} | e_k \rangle\!\approx\! 0.1$) character, as $E_z$ varies from $0$ to $35\,$mV/nm. In the same range of electric-field values, the excited states ($k=2,3$) change their symmetry along the $x$ direction, going from symmetric ($ \langle e_k | \hat\sigma_{zx} | e_k \rangle\!\approx\! 1$) to antisymmetric ($\langle e_k | \hat\sigma_{xy} | e_k \rangle\!\approx\! -0.8$). Further details are provided in the Supp. Mat. \cite{SuppMat} (Fig.~S11). 

The changes in the eigenstate symmetries are paralleled by stark transitions in their band compositions. In particular, while the states of the ground doublet remain predominantly heavy-hole in character ($w_{k;3/2,3/2}\gtrsim 0.95$), those of the first-excited doublet undergo a transition from a predominant light-hole composition ($w_{k;3/2,1/2} \approx 0.8$) to a strong heavy-hole character ($w_{k;3/2,3/2}\approx 0.9$). Further details are provided in the Supp.~Mat.~\cite{SuppMat} (Fig.~S6). {\color{black} Overall, the results presented in this Subsection show that the bias field can change the character of the lowest hole excitation: from a predominantly band character (same symmetries as the ground doublet, different band occupations) at low bias, to a predominantly charge character at large bias (same band occupations as the ground doublet, opposite symmetries).}

\paragraph{Rabi frequencies.}
The transitions that occur in the states of the first-excited doublet and, to a minor extent, in those of the ground doublet are dramatically reflected in the values of the Rabi frequencies (Fig.~\ref{f06}). All the $f_{z;jk}$ [panels (a,b,c)] display pronounced maxima in the transition region, approximately between 5 and 15 mV/nm, with values that are $1 \div 2$ orders of magnitude larger than those obtained for $E_z=50\,$mV/nm (see Fig.~\ref{f03} at $\theta = 75^\circ$ for a comparison). 

The frequencies $f_{x;jk}$ [panels (c,d,e)] undergo a transition in the same range of values of $E_z$. This leads to a pronounced increase of the Rabi frequencies, followed by a saturation at larger values of the bias field $E_z$. These transition amplitudes, therefore, are not maximized at the transition, where the excited states $|e_2\rangle$ and $|e_3\rangle$ change their symmetries, but in the high-field region ($E_z \gtrsim 15\,$mV/nm), where these eigenstates display well-defined symmetries along the oscillating-field direction ($x$). In the zero-bias limit ($E_z=0$), where all the considered eigenstates share the same symmetries along the $x$ direction, the Rabi frequencies $f_{x;jk}$ are strongly suppressed.

The above results confirm the connection between the Rabi frequencies and the relevant mirror symmetries of the eigenstates, which already emerged in Subsec.~\ref{subsec:proto}. The largest Rabi frequencies, namely those corresponding to interdoublet transitions, systematically increase as the two eigenstates involved in the transitions acquire well-defined and opposite symmetries along the direction of $\delta\boldsymbol{E}$. Local maxima are obtained in regions where the eigenstates undergo a transition, also reflected in their mirror symmetries, through the occurrence of minima in the expectation values of the operators $\hat\sigma_{\alpha\beta}$.

\subsection{Effect of uniaxial compressive strain\label{subsec:vs}}
\label{subsec:strain}

\begin{figure*}
\centering
\includegraphics[width=\textwidth]{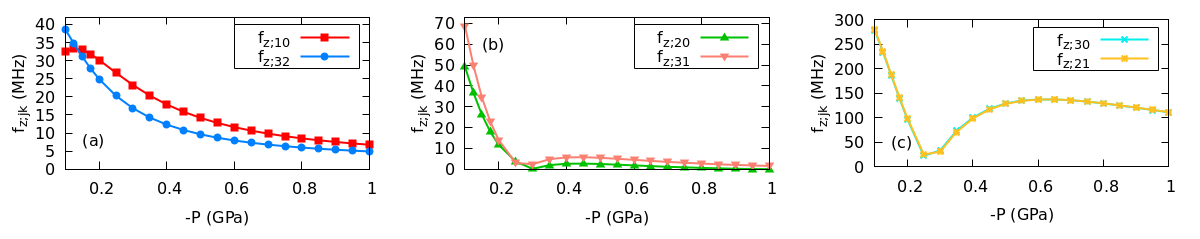}
\includegraphics[width=\textwidth]{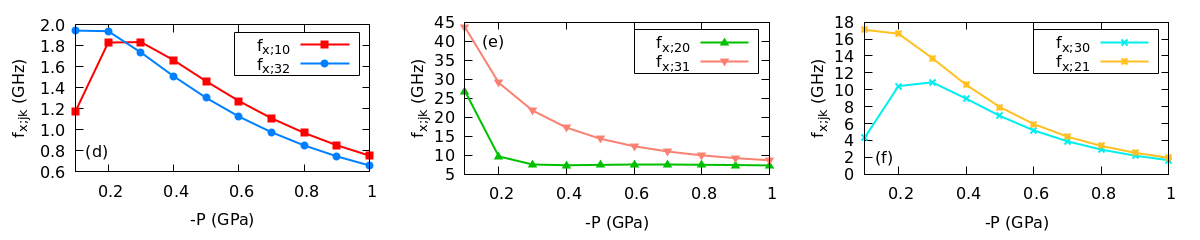}
\caption{Rabi frequencies (a,b,c) $f_{z;jk}$ and (d,e,f) $f_{x;jk}$ for QD1 as a function of the uniaxial compressive strain, associated to a time-dependent field of amplitude $\delta E=1\,$mV/nm, oriented along the $z$ and $x$ directions, respectively.  {\color{black} The values of $\omega_x$, $\omega_y$, $L_z$, $E_z$, $\theta$, and $B$ used in these calculations are reported in Table \ref{table1} (case C5) and in the table caption.}}
\label{f07}
\end{figure*}
Strain represents one of the possible means for engineering the properties of the spin-hole qubits \cite{Adelsberger22a,Abadillo23a,Secchi24a}. Tensile strain, in general, is known to reduce the amount of band mixing and the Rabi frequency within the ground doublet. Here, instead, we consider the effect of compressive strain{\color{black}, which is routinely used in pMOSFETs in order to enhance the hole mobility. The considered intensity and orientation of the uniaxial strain with respect to the crystallographic axes are also in line with those implemented in pMOSFETs \cite{Sverdlov}}. In particular, we report the dependence of the $f_{\alpha;jk}$ in QD1, as a function of the (negative) stress magnitude $P$ and for a fixed orientation of the applied magnetic field (case C5 in Table \ref{table1}).

{\color{black} The considered strain is oriented along the $x$ axis, which corresponds to the crystallographic direction [110] \cite{Sverdlov}. The effect of strain is accounted for by the Bir-Pikus Hamiltonian \cite{BirPikus}, whose expression is reported in the Supplemental Material, along with that of the relevant constants and of the strain tensor\cite{SuppMat}.}

\paragraph{Hole eigenstates.}
The application of a compressive uniaxial stress strongly affects the properties of the four lowest eigenstates, especially those belonging to the first excited doublet ($k =2,3$). In fact, even for modest values of the stress magnitude $P$, of the order of 10 MPa, the character of the first-excited doublet changes from that of an excitation along the $x$ direction, identified by 
\begin{gather}
\langle e_{k} | \hat{\sigma}_{yz} | e_k \rangle \approx - \langle e_k | \hat{\sigma}_{zx} | e_k \rangle \approx - 1\,,
\end{gather}
to that of an excitation along the $y$ direction, identified by 
\begin{gather}
\langle e_k | \hat{\sigma}_{yz} | e_k \rangle \approx - \langle e_k | \hat{\sigma}_{zx} | e_k \rangle \approx 1
\end{gather}
(see also Fig.~S12 in the Supp.~Mat.~\cite{SuppMat}). 

On a larger scale of stress-magnitude values, strain also produces an enhancement of the band mixing, with $w_{k;3/2,1/2}$ and $w_{k;1/2,1/2}$ reaching values above $0.15$ at $P=-1\,$GPa for all four hole eigenstates (see also Fig.~S7 in the Supp.~Mat.~\cite{SuppMat}). 

Overall, the application of compressive uniaxial stress abruptly changes the weak-confinement direction (from $x$ to $y$), and progressively increases the amount of band mixing.

\paragraph{Rabi frequencies.}
Strain produces different effects on the considered transition amplitudes.
The Rabi frequency $f_{z;21}$ increases by an order of magnitude for moderate values of the stress ($ 0 \lesssim -P \lesssim 0.2\,$GPa); in the same region, other frequencies either increase or are unaffected [see Fig.~\ref{f07}(a,b,c), and Fig.~\ref{f03} at $\theta = 75^\circ$ for a comparison]. Different behaviors are obtained also for larger values of the stress, but overall the $f_{z;jk}$ decrease for increasing values of $-P$. 

The effect of strain on the Rabi frequencies corresponding to $\delta\boldsymbol{E} \parallel \boldsymbol{u}_x$ [panels (d,e,f)] consists in a significant and systematic decrease, with respect to the values obtained for QD1 in the absence of stress. This is in line with the interpretation emerged so far, according to which the largest values of the Rabi frequencies are obtained when the oscillating field is oriented along the weak-confinement direction. As the applied stress modifies such direction (from $x$ to $y$), the Rabi frequencies associated to interdoublet transitions undergo a reduction by two orders of magnitude. {\color{black} This interpretation is confirmed by the significant enhancement of the Rabi frequencies $f_{y;jk}$ around $|P|\approx 0.1\,$GPa. However, the application of compressive strain along the weak-confinement direction is overall detrimental to the hole-state manipulation.}

{\color{black}
\section{Possible implications for quantum-information processing}
\label{sec: manipulation}

The large amplitudes of transitions that involve excited states can open up different possibilities for the hole-based encoding and manipulation of quantum information (Fig. \ref{f08}). These include: the use of an excited-doublet level for the encoding of the qubit states ``1", and the implementation of qubit-state flips through interdoublet transitions [panel (b)]; the use of ground- and excited-doublet states for the encoding of a qudit, where both intra- and inter-doublet transitions are required for the manipulation [panel (c)]; the conventional qubit encoding, based on the states of the ground doublet, combined with virtual inter-doublet transitions for the implementation of the qubit rotations [panel (d)]. In the latter case, the states belonging to the excited doublet(s) only serve as auxiliary levels, and are not meant to be populated throughout the qubit manipulation. Each of these approaches implies advantages and disadvantages, whose detailed analysis will require further investigation. In the present Section, we provide a preliminary discussion of the main aspects and figures of merit.

\subsection{Manipulation and readout}

In the first two approaches mentioned above [Fig.~\ref{f08}(b-c)], as well as in the usual approach [panel (a)], the logical operations are directly related to the transitions between eigenstates. One of the possible challenges is represented by the need of spectrally resolving transitions with close energies $\Delta_{jk}$. As shown in Fig.~\ref{f02}, for the considered confinement and magnetic-field values, the differences between intra- and inter-doublet gaps are of the order $0.1$ and $1$ meV, respectively. These values are large compared to transition amplitudes of the order of $1\,{\rm GHz} \approx 4.14\,\mu$eV, such as the ones obtained for the $f_{z;jk}$, but are small compared to the THz range obtained for the $f_{x;jk}$. In this latter case, being the Rabi frequency comparable to the resonant frequency of the transition, one would in fact enter the ultrastrong coupling regime, which opens up the possibility of implementing ultrafast quantum computing \cite{FornDiaz19a}. In cases where only one interdoublet transition is required for the manipulation [Fig.~\ref{f08}(b)], the different dependence of the Rabi frequencies on tunable parameters, such as the magnetic field orientation and the intensity of the bias field (see Section \ref{sec: results}), can be exploited to maximize the ratio between the amplitudes of the required transitions and those of the detrimental ones. The implementation of genuinely multilevel manipulation schemes [Fig. \ref{f08}(c)] would instead require the use of optimal-control approaches \cite{Koch22a}. 

\begin{figure}[t]
\centering
\includegraphics[width=0.45\textwidth]{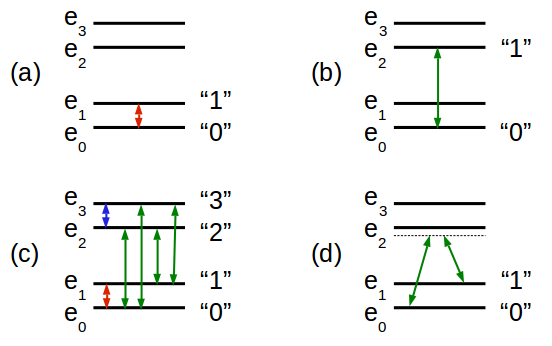}
\caption{\color{black} Possible encodings based on the use of the hole states: (a) usual qubit encoding, where the transition within the ground doublet is used for the manipulation; (b) alternative qubit encoding, where the qubit state ``1" is identified with an excited doublet state; (c) qudit encoding, where each of the four hole levels is identified with a logical state; (d) usual qubit encoding, where however the manipulation exploits virtual transitions to the excited doublet states.}
\label{f08}
\end{figure}

In the last of the mentioned approaches [Fig. \ref{f08}(d)], the qubit states are coupled by Raman transitions, whose amplitudes can be derived from those of the involved interdoublet transitions. In the reported example the amplitude of the Raman transition would be given by $ f_{\alpha;01}^{\rm eff} \equiv h\,f_{\alpha;02} \, f_{\alpha;12} / |\Delta| $, being $\Delta$ the common detuning of the fields inducing the two transitions \cite{Gardiner00a}. The inequality $|\Delta |/ h \gg f_{\alpha;02} , f_{\alpha;12}$ defines the virtual Raman regime, where the actual occupation of the interconnecting excited state $|e_2\rangle$ is kept negligible throughout the process. Given the difference of several orders of magnitude between $f_{\alpha;01}$ and the Rabi frequencies of the interdoublet transitions, the above inequality still allows the achievement of a transition amplitude that is large, compared to that of the direct transition: $f_{\alpha;01}^{\rm eff} \gg f_{\alpha;01}$. While the above expression of $f_{\alpha;01}^{\rm eff}$ provides a first estimate of the Raman transition amplitude, a more accurate calculation might require the inclusion of more contributions, related to diverse interconnecting excited states.
 
As to the readout, we note that one of the possible approaches [Fig.~\ref{f08}(d)] is based on the standard qubit encoding, and implies no further requirements. For the alternative qubit encoding [Fig.~\ref{f08}(b)], one might envision a mapping of the excited state $|1\rangle \equiv |e_2\rangle$ onto the state $|e_1\rangle$ at the end of the quantum-gate sequence, followed by the usual readout process. Generalizations of the readout scheme to the case of qudits [Fig.~\ref{f08}(c)] would instead require nontrivial generalizations of the current readout approaches.

We finally note that the transitions involving excited hole states unavoidably play a role in possible multihole qubit encodings. Let us call $|\Psi_0\rangle$ and $|\Psi_1\rangle$ the two states belonging to the ground doublet of a $(2n+1)$-hole system. Each of these multihole states will generally consist of a linear superposition of different Slater determinants $|\Phi_i\rangle$, each one corresponding to the occupation of a set of $2n+1$ single-hole levels: $|\Psi_k\rangle = \sum_i C_{i,k} |\Phi_i\rangle$. The expectation value of the electric dipole operator, which enters the expression of the Rabi frequencies, is thus given by: $ \langle\Psi_0 | {\bf r} |\Psi_1\rangle = \sum_{i} |C_{0,i}|^2 \langle\Phi_i | {\bf r} |\Psi_i \rangle + \sum_{i \neq j} C_{0,i}^* C_{1,j} \langle\Phi_i | {\bf r} |\Phi_j \rangle $. The relevant terms in the second sum are those involving Slater determinants that differ by one single-hole level, all the others being identically zero. If we call $|e_{k_i}\rangle$ ($|e_{k_j}\rangle$) the single-hole state that is occupied in $|\Phi_i\rangle$ ($|\Phi_j\rangle$) and unoccupied in $|\Phi_j\rangle$ ($|\Phi_i\rangle$), then one has that:
$\langle\Phi_i | {\bf r} |\Psi_j \rangle = \langle e_{l_i} | {\bf r} |e_{l_j} \rangle$. From this it follows that the Rabi frequency in multi-hole qubits is also determined by the amplitudes of transition involving single-hole excited states. In particular, if these can be orders of magnitude larger than those between $|e_0\rangle$ and $|e_1\rangle$, as the present results show, then even limited occupations of excited levels $|e_{i>1}\rangle$ in the eigenstates $|\Psi_j\rangle$ can provide significant contributions to the overall Rabi frequency of the multi-hole qubit.

\subsection{Decoherence}

One of the possible approaches based on the use of interdoublet transitions [Fig.~\ref{f08}(d)] would not be affected by the decoherence time $\tau$ of the excited states, because in the virtual Raman regime the interconnecting levels are not significantly populated, provided that $h/\tau < |\Delta|$. If instead the excited states encode information, then the qubit or qudit manipulation implies their population, and the gate fidelities will be affected by their decoherence times, which might be shorter than those characterizing ground doublet. This expectation is based on the consideration that the state $|e_1\rangle$ is regarded as a ``spin excitation" (with respect to the ground state $|e_0\rangle$), whereas $|e_2\rangle$ and $|e_3\rangle$ would be ``charge excitations". We note that this expectation is partially correct. In fact, as the discussion reported in the previous Section shows, the distinction between charge and spin -- actually, band -- excitations is not always clear, because both components can be found in the states $|e_2\rangle$ and $|e_3\rangle$. 

\begin{figure}[t]
\centering
\includegraphics[width=0.45\textwidth]{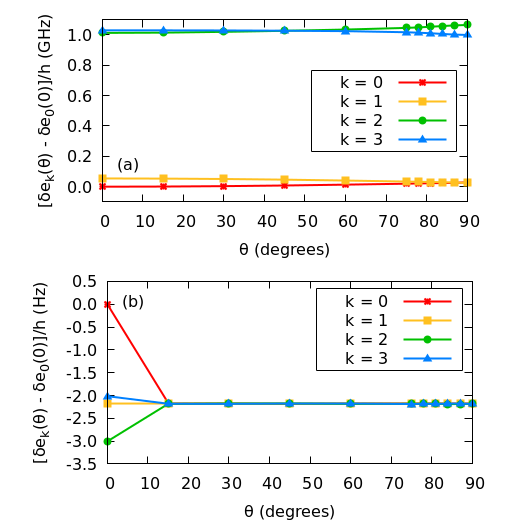}
\caption{\color{black} Dependence of the hole energies $e_k$ on an electric field $E_{\rm cn}=1\,$mV/nm oriented along the (a) $z$ and (b) $x$ directions. The dot parameters are those reported in Table \ref{table1} for QD1, case C1.}
\label{f09}
\end{figure}

As in the case of the ground doublet, the main sources of decoherence for the excited states can be identified with hyperfine interactions, phonons, and charge noise \cite{Burkard21a,Fang23}. In particular, charge noise can induce fluctuations in the energy gaps $\Delta_{jk}$, resulting in inhomogeneous dephasing on a characteristic time scale $T_2^*$. As a preliminary analysis of the problem, we consider the dependence of the hole energies $e_k$ on a charge-noise related electric field $ E_{\rm cn}$ oriented along the $x$ or $z$ directions (Fig.~\ref{f09}). 

If the fluctuating field is oriented along the $z$ direction [panel (a)], two states belonging to the same doublet (ground or excited) undergo energy shifts $\delta e_k = \boldsymbol{E}_{\rm cn} \cdot \langle e_k | \boldsymbol{\hat r}| e_k \rangle$ whose difference is much smaller than that between states of belonging to different doublets:
$ |\delta e_1 - \delta e_0 | \approx |\delta e_3 - \delta e_2 | \ll |\delta e_j - \delta e_k |$,
with $j=0,1$ and $k=2,3$. This implies that a linear superposition between, e.g., $|e_0\rangle$ and $|e_2\rangle$ would indeed be affected by a much faster dephasing than one between $|e_0\rangle$ and $|e_1\rangle$, or $|e_2\rangle$ and $|e_3\rangle$.

A qualitatively different behavior is found for a charge-noise field oriented along the $x$ direction [panel (b)]. In this case, the energy fluctuations $\delta e_k$ are nine orders of magnitude smaller than for $\boldsymbol{E}_{\rm cn} \parallel \boldsymbol{u}_z$. Moreover, there is no substantial between states belonging to the same or to different doublets. In fact, for tilting angles $\theta \gtrsim 15^\circ$, all the energy shifts $\delta e_k$ take very similar values, so that a fluctuating electric field would not induce significant variations in the energy gaps. 

From this preliminary analysis it emerges that inhomogeneous dephasing affecting linear superpositions of ground and excited doublet states strongly depends on the orientation of the fluctuating electric field. As for the Rabi frequencies $f_{x;jk}$, the suppression of the $\delta e_k$ for $\boldsymbol{E}_{\rm cn} \parallel \boldsymbol{u}_x$ can be related to symmetry, and specifically to the fact that states $|e_k\rangle$ with a well-defined mirror symmetry along the $x$ direction cannot have a significant expectation value of a $\hat{\sigma}_{yz}$-odd operator, such as $\hat{x}$.
}

\section{Conclusions}
\label{sec: conclusions}

So far, a great deal of attention has been devoted to the electrical manipulation of the hole state within the ground doublet. Transitions involving higher hole states are also of potential interest. On the one hand, they might become relevant for alternative schemes, which involve excited levels in the encoding and manipulation of quantum information. On the other hand, in multi-hole implementations of the spin qubit, Rabi frequencies involving excited states unavoidably contribute to the transition amplitudes between few-particle multi-configurational states. 

Several robust conclusions can be drawn from the whole set of numerical results, focused on the two lowest doublets of a slightly elongated quantum dot. First, the Rabi frequencies involving the excited states generally are one or two orders of magnitude larger than that involving the ground-doublet states, typically used for the qubit encoding. Second, three or four orders of magnitude can be further gained by applying the oscillating electric field in the plane orthogonal to the growth direction, and specifically in the weakest confinement direction, along which the ground and excited states can be tuned so as to display well-defined and opposite symmetries. Third, in the absence of such a condition, pronounced maxima in the Rabi frequencies are obtained in regions of the parameter space where transitions in the character of the eigenstates take place. Similar transitions are reflected in the occurrence of avoided level crossings or of minima in the expectation values of the mirror symmetry operators. 

\acknowledgments
The authors acknowledge financial support from the Ministero dell’Universit\`a e della Ricerca (MUR) under the Project PRIN 2022 number 2022W9W423 and the PNRR Project PE0000023-NQSTI.  


%

\end{document}